\documentclass[journal]{IEEEtran}
\ifCLASSINFOpdf
   \usepackage[pdftex]{graphicx}
\else
   \usepackage[dvips]{graphicx}
\fi

\usepackage[utf8]{inputenc}

\usepackage{xcolor}
\usepackage{color, soul}
\usepackage{enumerate}
\usepackage[shortlabels]{enumitem}
\usepackage{url}

\usepackage{hyperref}
\usepackage{cite}
\usepackage{bm, bbm} 
\usepackage{mathtools}
\usepackage{amsmath, amssymb, amsthm}
\usepackage{nicefrac}
\usepackage{empheq}

\usepackage{booktabs}
\usepackage{graphicx}
\usepackage{multirow}
\usepackage{units}
\usepackage{siunitx}
\usepackage{enumitem}

\usepackage{placeins}
\newcommand{\subparagraph}{}

\usepackage{amsmath,amsfonts,amsthm,amssymb}
\usepackage[bb=boondox]{mathalfa}
\usepackage[algoruled]{algorithm2e}
\usepackage{array}

\usepackage[noabbrev,capitalize]{cleveref}

\crefname{equation}{}{}
\Crefname{equation}{}{}

\theoremstyle{definition} % bold title, normal text

\theoremstyle{plain} % bold title, italic text

\theoremstyle{remark} % italic title, normal text

\usepackage{tikz}

\newcommand{\set}[1]{\mathcal{#1}} % for caligraphed set-symbols
\makeatletter
\newcommand{\pushright}[1]{\ifmeasuring@#1\else\omit\hfill$\displaystyle#1$\fi\ignorespaces}
\newcommand{\pushleft}[1]{\ifmeasuring@#1\else\omit$\displaystyle#1$\hfill\fi\ignorespaces}
\makeatother

\newcolumntype{C}[1]{>{\centering\arraybackslash}p{#1}} % for columns with auto-line break like p{} but centered

\makeatletter
\let\old@ps@headings\ps@headings
\let\old@ps@IEEEtitlepagestyle\ps@IEEEtitlepagestyle
\def\psccfooter#1{%
    \def\ps@headings{%
        \old@ps@headings%
        \def\@oddfoot{\strut\hfill#1\hfill\strut}%
        \def\@evenfoot{\strut\hfill#1\hfill\strut}%
    }%
    \def\ps@IEEEtitlepagestyle{%
        \old@ps@IEEEtitlepagestyle%
        \def\@oddfoot{\strut\hfill#1\hfill\strut}%
        \def\@evenfoot{\strut\hfill#1\hfill\strut}%
    }%
    \ps@headings%
}
\makeatother

\begin{document}

\title{Operation-Adversarial Scenario Generation}

\author{
\IEEEauthorblockN{Zhirui Liang, Robert Mieth, Yury Dvorkin}
\IEEEauthorblockA{Department of Electrical and Computer Engineering, New York University, USA\\
\{zl3364, robert.mieth, dvorkin\}@nyu.edu
}
}

\maketitle

\begin{abstract}
This paper proposes a modified conditional generative adversarial network (cGAN) model to generate net load scenarios for power systems that are statistically credible, conditioned by given labels (e.g., seasons), and, at the same time, ``stressful'' to the system operations and dispatch decisions. The measure of stress used in this paper is based on the operating cost increases due to net load changes. 
The proposed operation-adversarial cGAN (OA-cGAN) internalizes a DC optimal power flow model and seeks to maximize the operating cost and achieve a worst-case data generation. The training and testing stages employed in the proposed OA-cGAN use historical day-ahead net load forecast errors and has been implemented for the realistic NYISO 11-zone system. Our numerical experiments demonstrate that the generated operation-adversarial forecast errors lead to more cost-effective and reliable dispatch decisions.
\end{abstract}

% Use this to place sponsorships
% \thanksto{\noindent Submitted to the 22nd Power Systems Computation Conference (PSCC 2022).}

\vspace{-0.3cm}
\section{Introduction}
\label{sec:Introduction}

Dealing with renewable energy sources requires internalizing their stochasticity into optimization and market-clearing tools used in power system operations. 
To this end, stochastic \cite{zakaria2020uncertainty} and robust \cite{nazari2018application} optimization methods have been employed and demonstrated to improve power system cost efficiency and reliability. 
Consider a decision-making problem, such as an optimal power flow or unit commitment problem $\min_{\substack{x\in\set{X}(\omega)}} {C}(x,\omega)$, where $C(\cdot)$ and $x$ are the objective (cost) function and the vector of decision variables (e.g. generator outputs) constrained by feasible solution space $\set{X}$ and $\omega \in \Omega$ is a vector of uncertain parameters (e.g. net load or renewable injections) affecting both the objective function and the feasible region. 
In stochastic approaches, this problem is often solved by representing $\Omega$ as a set of discrete scenarios $\{\omega_s\}$ with probability $\pi_s$ and minimizing the expected cost across all scenarios, i.e., $\min_{\substack{x\in\set{X}(\omega_s)}} \sum_s \pi_s C(x,\omega_s)$.
However, besides high computational requirements that limit the number of scenarios that can be considered, the accuracy of the scenario-based method highly depends on how well the chosen scenarios can capture both the range and  correlation structures of  uncertain parameters. 
Ideally, the used scenarios are historical realizations of  uncertain parameters, which will ensure the most accurate representation. 
However, relying on historical samples raises two  challenges. 
First, the number of historical samples may be scarce. For example, if a power system operator wants to analyze the impact of a wind farm that is under construction, there are no historical data points available to use as scenarios. Thus, scenarios with credible statistical properties need to be \textit{synthesized}.  
Second, historical samples may not include events that are relevant in the future, i.e., considering the most adversarial historic event to ensure system reliability may not be sufficient if a new and worse event materializes. 

The first challenge can be addressed through novel data-driven approaches that are powerful in catering to specific requirements based on the underlying patterns they learned from historical data. 
Specifically, a machine learning framework named \textit{generative adversarial network} (GAN) proposed by Goodfellow \textit{et al.} \cite{goodfellow2014generative} has been shown to efficiently synthesize data samples that fit to a given empirical distribution with very high credibility.
Further, a modification of GANs introduced by Mirza and Osindero \cite{mirza2014conditional}, called conditional GAN (cGAN), allows to \textit{condition} the generated data sets based on predefined labels, thus allowing to tune the generated synthetic scenarios to the specific needs of their applications.  
GANs and cGANs have been successfully applied to power system problems. 
For example, Chen \textit{et al.} \cite{chen2018model} used cGAN to generate scenarios for wind and solar injections. 
Further power system applications of (c)GANs include Wang \textit{et al.} \cite{wang2020modeling} who applied cGANs to generate load scenarios, Zhang \textit{et al.} \cite{zhang2020typical} who studied wind power injection scenarios with focus on the spatio-temporal correlation between multiple wind farms in the system, and Wang \textit{et al.} \cite{wang2019generative} who used GANs to improve short-term forecasting of renewable injections. 
While the approaches in \cite{chen2018model,wang2020modeling,zhang2020typical,wang2019generative} can successfully synthesize the statistical properties of the historical data they do \textit{not} consider the impact of a  (c)GAN-generated data sample on the decision making problem $\min_{\substack{x\in\set{X}(\omega)}} {C}(x,\omega)$ at hand, thus they do not address the second challenge of scenario generation. 
Traditionally, this challenge has been addressed by generating robust scenarios that may not be statistically credible but constitute a worst-case outcome for the decision-making task, i.e., $  \min_{\substack{x\in\set{X}(\omega)}} \sup_{\substack{\omega \in \set{U}}} {C}(x,\omega)$, where $\set{U}$ is a predefined uncertainty set.  
Such robust decisions are usually overly conservative and, therefore, costly. Additionally, the analytic and/or computationally tractable solutions to the inner maximization problem may be difficult to obtain for some $\set{U}$ and, thus, often require approximations that further add to the solution conservatism. 

To address these two scenario generation challenges simultaneously, we propose a modified cGAN that can generate scenarios that are adversarial for the decision-making task, but, at the same time, remain statistically credible. We summarize the contributions of this paper as follows: 
\begin{itemize}
    \item Unlike the previous work in \cite{chen2018model,wang2020modeling,zhang2020typical,wang2019generative}, we internalize the decision-making task, in our case a DC optimal power flow (DC-OPF) problem, into the cGAN training phase, thus rendering it \textit{operation adversarial} (OA-cGAN).
    \item We use the proposed OA-cGAN to generate worst-case forecast errors of the real-time net load (i.e., demand minus renewable injections) during a day-ahead planning stage, which is one of the possible application scenarios of the proposed OA-cGAN framework. Our implementation and experiments use real-world data from the New York Independent System Operator (NYISO). 
    \item We derive the necessary training method and demonstrate that the proposed OA-cGAN generates statistically credible forecast errors that inform robust and cost effective reserve allocation decisions.
\end{itemize}

\section{Preliminaries}
\label{sec:Basic_model}

\subsection{Conditional generative adversarial networks (cGANs)}
\label{subsec:cGANs}

\begin{figure}[!t]
    \centering
    \includegraphics[width=1\linewidth]{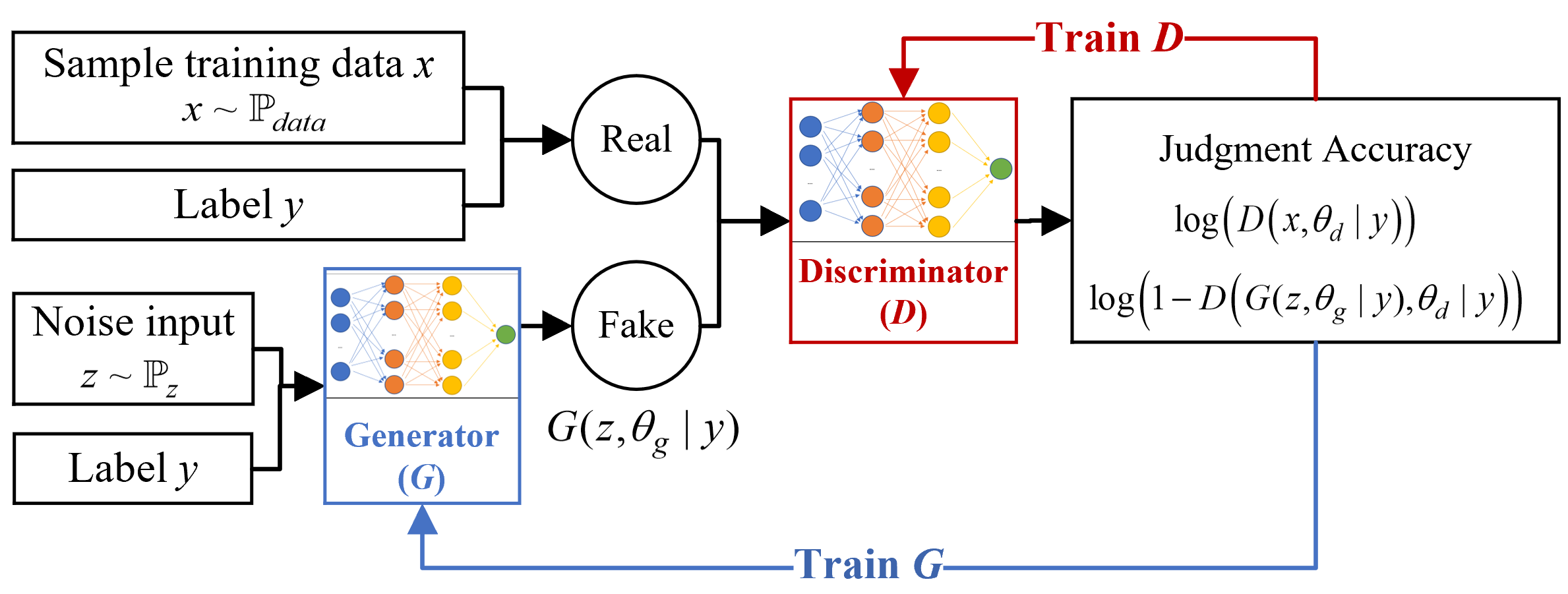}
    \caption{A typical structure of the cGAN model.}
    \label{fig:cGAN}
\end{figure}

Fig.~\ref{fig:cGAN} illustrates a basic cGAN model with one generator ($G$) and one discriminator ($D$), both of which are non-linear mapping functions, such as neural networks. Generator $G$ and discriminator $D$ are defined by a set of parameters $\theta_g$ and $\theta_d$, respectively, which must be trained. Specifically, $G$ and $D$ are trained alternately in a zero-sum game process. The training objective of $G$ is to tune $\theta_g$ such that $G$ transforms data samples drawn from some distribution $\mathbb{P}_z$ into new data points that follow a target data distribution $\mathbb{P}_{data}$. 
The training objective of  $D$, on the other hand, is to tune $\theta_d$ such that $D$ can distinguish real data samples drawn from $\mathbb{P}_{data}$ and synthetic data points generated by $G$ with high accuracy.
The additional input is label $y$, which conditions the training of $G$ and $D$ to specific data features, hence the naming convention ``conditional'' GAN.
The \textit{adversarial} competition between the objectives of $G$ and $D$ will push each model to improve its performance until a Nash equilibrium is reached, i.e., the samples produced by $G$ cannot be distinguished from the original data by $D$. 
The full training objective function of cGANs can then be formalized as:
\begin{align}
    \min_{\substack{\theta_g}}\max_{\substack{\theta_d}}\ &\mathbb{E}_{x \sim \mathbb{P}_{data}}[\log \left( D(x,{\theta_d}|y) \right)] \nonumber \\ 
    &+ \mathbb{E}_{z \sim \mathbb{P}_z}[\log \left(1 - D(G(z,{\theta_g}|y),{\theta_d}|y)\right)],  \label{cGAN}
\end{align}
where $x\sim{\mathbb{P}_{data}}$ is data from the real distribution, $z\sim{\mathbb{P}_z}$ is randomly generated data (e.g. from a Gaussian distribution), $G(z,{\theta_g}|y)$ is the output of $G$, i.e., the generated data based on the noise input (denoted as $z$) and label $y$, and $D(x,{\theta_d}|y)$ is the output of $D$, i.e., the probability that $x$ is from real data distribution $\mathbb{P}_{data}$ conditioned by label $y$ ($D(x,{\theta_d}|y) \in [0,1]$).
Operators $\mathbb{E}_{x \sim \mathbb{P}_{data}}$ and $\mathbb{E}_{z \sim \mathbb{P}_z}$ compute the expectation with respect to distributions $\mathbb{P}_{data}$ and $\mathbb{P}_z$, respectively.

Training objective \eqref{cGAN} is achieved by alternately tuning $\theta_g$ such that $G$ maximizes the probability that the currently trained $D$ identifies its synthetic data $G(z,{\theta_g}|y)$ as real:
\begin{align}
    \max_{\substack{\theta_g}}\ \mathbb{E}_{z \sim \mathbb{P}_z}\left[ {\log \left( {D(G(z,{\theta_g}|y), {\theta_d}|y)} \right)} \right], \label{GAN_G}
\end{align}%
and tuning $\theta_d$ such that $D$ maximizes its judgement accuracy, i.e., achieving high values $D(x,{\theta_d}|y)$ for real data and low values $D(G(z,{\theta_g}|y),{\theta_d}|y)$ for synthetic data:
\begin{align}
    \max_{\substack{\theta_d}}\ &\mathbb{E}_{x \sim \mathbb{P}_{data}}\left[ {\log \left(D(x,{\theta_d}|y) \right)} \right]  \nonumber \\  &+\mathbb{E}_{z \sim \mathbb{P}_z}\left[ {\log \left( {1-D(G(z,{\theta_g}|y), {\theta_d}|y)} \right)} \right]. \label{GAN_D}
\end{align}%

\subsection{Power system operation model}
\label{subsec:DCOPF}

We consider a standard DC optimal power flow (DC-OPF) problem to model power system operations. The DC-OPF minimizes the operating cost of supplying the system net load (i.e., load minus renewable injections) with respect to physical limits of generators and transmission lines:
\allowdisplaybreaks
\begin{subequations}
\begin{align}
&\min_{\substack{\{P_{g,t}\}_{g\in \set{G}, t\in \set{T}},\\ \{\theta_{i,t}\}_{i\in \set{I}, t\in \set{T}}}}\ 
    C = \sum\limits_{t\in \set{T}} \sum\limits_{g\in \set{G}} {(c_{0g} + c_{1g}P_{g,t} + c_{2g}P_{g,t}^2)} \label{DCOPF_objective}\\
    & \text{s.t. } \nonumber \\
    &(\lambda_{i,t}):\sum\nolimits_{g\in \set{G}_i}P_{g,t} - \sum\nolimits_{j \in \mathcal{N}_i } B_{i,j} (\theta_{i,t} - \theta_{j,t}) = d_{i,t} \nonumber \\
    & \hspace{1.1cm}   
     \forall{i}\in\set{I},\ \forall{t}\in\set{T} \label{DCOPF_power_balance}\\
    &(\rho_{g,t}^{-},\rho_{g,t}^{+}): 0 \le P_{g,t} \le P_{g}^{\max} \quad \forall{g}\in\set{G} \label{DCOPF_Pmax}\\
    & (\beta_{i,j,t}^{-},\beta_{i,j,t}^{+}): -S_{i,j} \le B_{i,j} (\theta_{i,t}-\theta_{j,t}) \le S_{i,j}  \nonumber \\ 
    & \hspace{1cm}\forall{i}\in\set{I},\ \forall{j}\in \mathcal{N}_i,\ \forall{t}\in\set{T}  \label{DCOPF_power_flow}\\
    & (\eta_t): \theta_{ref,t} = 0 \quad \forall{t}\in\set{T}, \label{DCOPF_ref}
\end{align}%
\label{DCOPF}%
\end{subequations}%
\allowdisplaybreaks[0]%
where $\set{I}$ is the set of nodes in the transmission network indexed by $i$, $\set{T}$ is the set of time steps in the planing horizon indexed by $t$, $d_{i,t}$ is the net load at node $i$ and time $t$, $P_{g,t}$ is the (active) power output of generator $g$ at time $t$, $\set{G}_i$ is the set of generators connected to node $i$, $\mathcal{N}_i $ is the set of nodes adjacent to $i$, $\theta_{i,t}$ is the voltage angle at node $i$ at time $t$, $B_{i,j}$ is the susceptance of the line between node $i$ and $j$, and $S_{i,j}$ is the thermal capacity of the line between node $i$ and $j$. 
Objective~\eqref{DCOPF_objective} minimizes system cost using a quadratic cost model of each generator given by parameters $c_{0g}$, $c_{1g}$, $c_{2g}$. 
Eq.~\cref{DCOPF_power_balance} enforces the nodal power balance at each node. Eqs.~\cref{DCOPF_Pmax} and \cref{DCOPF_power_flow} limit the output of generators and the power flow on each line to their technical limits. Eq.~\cref{DCOPF_ref} sets the voltage angle at the reference node ($i=ref$) to 0. 
Greek letters in parentheses in \cref{DCOPF_power_balance,DCOPF_Pmax,DCOPF_power_flow,DCOPF_ref} denote dual multipliers of the respective constraints.

\section{Operation-Adversarial cGAN model}
\label{sec:Proposed_model}

\subsection{Training objective}
\label{subsec:Objective}

\begin{figure}[!t]
    \centering
    \includegraphics[width=1\linewidth]{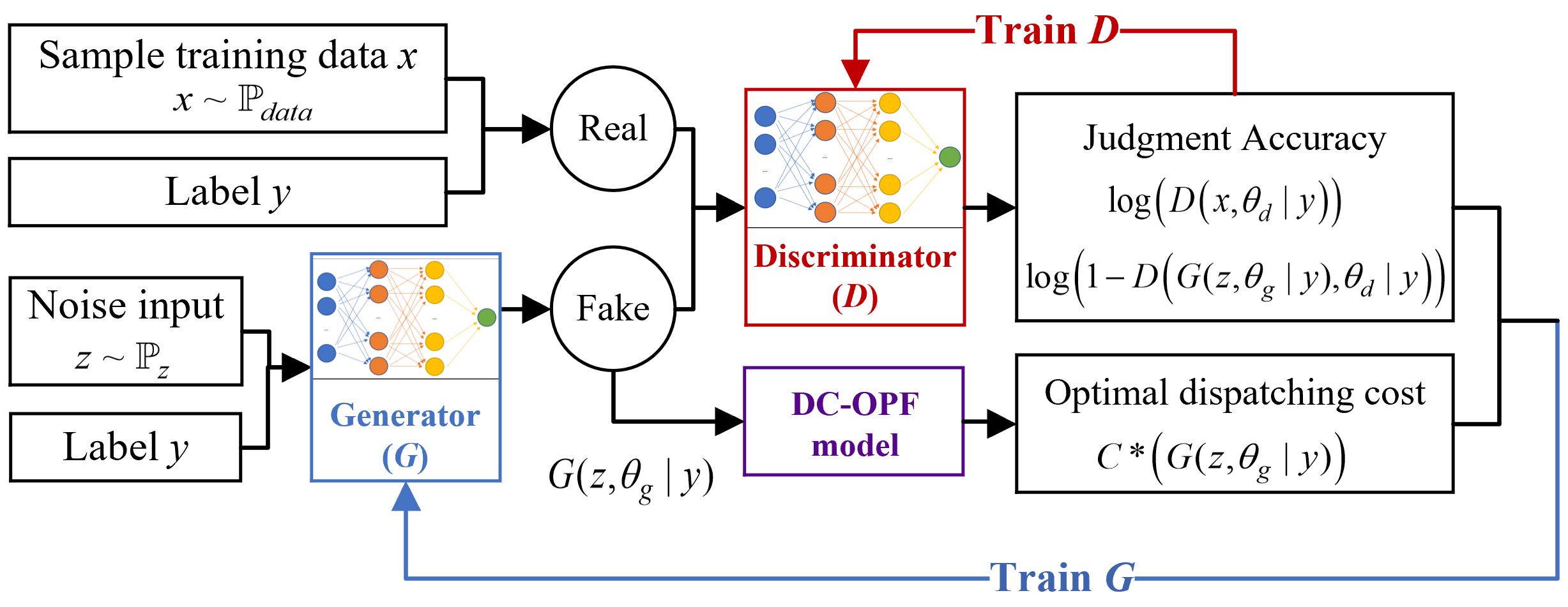}
    \caption{The proposed structure of the OA-cGAN model.}
    \label{fig:operation_adversarial_cGAN}
\end{figure}
The structure of the proposed operation-adversarial cGAN model (OA-cGAN) is shown in Fig.~\ref{fig:operation_adversarial_cGAN}. Compared with the traditional cGAN in Fig.~\ref{fig:cGAN}, another player (i.e., the DC-OPF model) joins the game between $G$ and $D$ and becomes part of the training process of $G$. Note that the training process of $D$ remains the same as in the traditional cGAN model. Thus, the objective of $D$ in the OA-cGAN model is still maximizing its judgment accuracy as shown in \cref{GAN_D}. 
As the first step, we formulate the training objectives of $G$ and $D$ as an equivalent minimization problem to achieve consistency with the cost-minimizing DC-OPF formulation:
\begin{align}
\min_{\substack{\theta_d}}\ loss_D &= \mathbb{E}_{x \sim \mathbb{P}_{data}}\left[ {\log \left(1-D(x,{\theta_d}|y) \right)} \right] \nonumber \\ &+\mathbb{E}_{z \sim \mathbb{P}_z}\left[ {\log \left( {D(G(z,{\theta_g}|y), {\theta_d}|y)} \right)} \right] \label{model_D}.
\end{align}%

Next, the training objective of generator $G$ receives an additional component to capture the operational model:
\begin{align}
\min_{\substack{\theta_g}}\
    loss_G =& k \overbrace{ \mathbb{E}_{z \sim \mathbb{P}_z}\left[ {\log \left( {1-D(G(z,{\theta_g}|y), {\theta_d}|y)} \right)} \right]}^{loss_{G1}} \nonumber \\
    & +(1-k) \underbrace{ \mathbb{E}_{z \sim \mathbb{P}_z}\left[ -{C^{*} \left( G(z,{\theta_g}|y) \right)} \right]}_{loss_{G2}}. \label{objective_G}
\end{align}%
The first part of \cref{objective_G} (denoted as $loss_{G1}$) maximizes the probability that the generated data is recognized as real data by $D$, i.e., playing against $D$ as in \eqref{GAN_G}, and the second part (denoted as $loss_{G2}$) maximizes the expected operating cost based on the generated net load, i.e., playing against the DC-OPF. 
The two objectives are weighted against each other using factor $k\in[0,1]$. When $k=1$, the OA-cGAN becomes a traditional cGAN. Term $C^{*} \left( G(z,{\theta_g}|y) \right)$ in \cref{objective_G} is interpreted as the scaled optimal operating cost based on  generated load $G(z,{\theta_g}|y)$, i.e.,:
\begin{align}
    C^*=  \frac {\sum\nolimits_{t\in \set{T}} \sum\nolimits_{g\in \set{G}} (c_{0g} + c_{1g}P_{g,t}^* + c_{2g}{P_{g,t}^*}^2)-\delta_{shift}} {\delta_{scale}} \label{C_star}
\end{align}
where $P_{g,t}^*$ is the optimal power output of generator $g$ at time $t$ obtained by solving the DC-OPF based on generated load $G(z,{\theta_g}|y)$. Since $loss_{G1}$ represents a \textit{probability} and therefore will always take values between 0 and 1, we use constants $\delta_{shift}$ and $\delta_{scale}$ to project operating the cost into a comparable interval. This will allow for trading off the two parts of the objective using weight $k$. 

\begin{figure}[!t]
    \centering
    \includegraphics[width=0.9\linewidth]{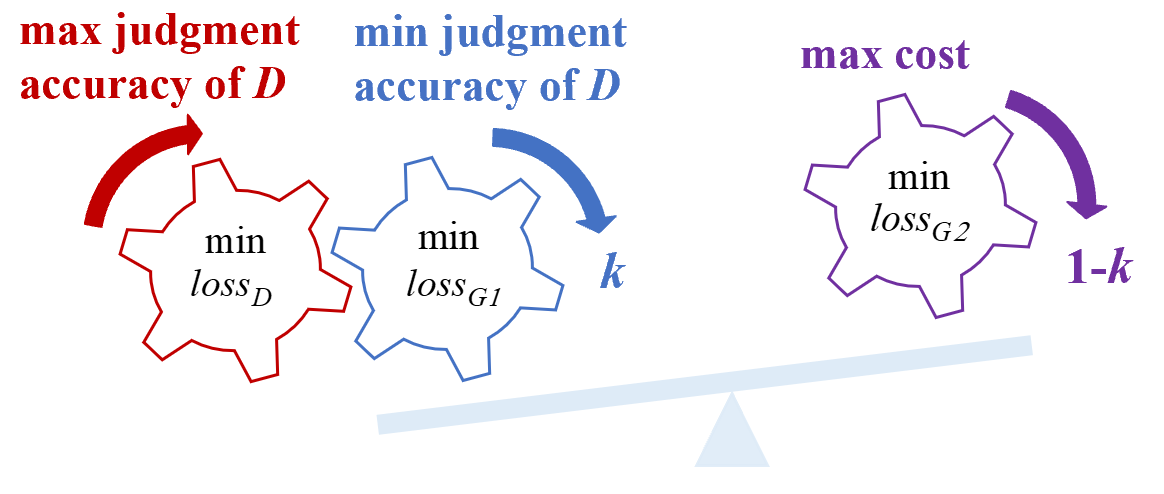}
    \caption{Relationships between three objectives in OA-cGAN.}
    \label{fig:three_players}
\end{figure}

Fig.~\ref{fig:three_players} illustrates the relationship between the three objectives ($\min \ loss_{D}$, $ \min \ loss_{G1}$, and $\min \ loss_{G2}$) of the OA-cGAN. First, objectives $loss_{G1}$ and $loss_{D}$ capture the competition between the credibility of scenario generation from $G$ and the detection accuracy of $D$. Second, objective $loss_{G1}$ and $loss_{G2}$ determine the success of $G$ to either work against $D$ or the DC-OPF, respectively. Depending on weight $k$, $G$ prioritizes the former or latter objective. 
Specifically, by focusing on minimizing $loss_{G1}$, the generated data becomes more statistically credible, while by minimizing $loss_{G2}$, the generated data become operational-adversarial. 

\subsection{Training data preparation}
\label{subsec:Training}
In this paper, we describe the training process for operation-adversarial scenarios on the \textit{net load forecasts}, i.e., the difference between the forecast net load during a day-ahead (DA) planning stage and the realized net load in real-time (RT). 
We note that the proposed OA-cGAN can be adapted to other scenario parameters, e.g., renewable injections. 
It is also assumed the planning horizon is one day and has a resolution of 1 hour, i.e., $\mathcal{T}=\{1,...,24\}$.

The training process requires a suitably prepared training data set, which we create as follows:

\noindent
\underline{\textit{Step 1}}: Obtain historical data for DA and RT net loads for each sample day $s$ in the training and testing data sets defined as $\{DA_s, RT_s\}_{\ s \in \set{S}_{train} \cup \set{S}_{test}}$, where $DA_{i,t,s}$ and $RT_{i,t,s}$ are DA load forecast and RT actual loads for each node $i\in\set{I}$ and $t\in\set{T}$ in sample day $s$. Each sample $s$ receives label $y_s$, which denotes attributes of interest such as the day of the week, month, season, and weather conditions in that day.

\noindent
\underline{\textit{Step 2}}:  For each $s$ and $i$, calculate the minimum, average and maximums denoted as $DA^{\min}_{i,s}$, $DA^{ave}_{i,s}$ and  $DA^{\max}_{i,s}$.

\noindent
\underline{\textit{Step 3}}: For each $s$, $i$ and $t$, calculate a normalized DA and RT load ($DA^{norm}_{i,t,s}$ and $RT^{norm}_{i,t,s}$) as:
\begin{align}
    DA^{norm}_{i,t,s} &= \left({DA_{i,t,s} {\rm{-}} DA^{ave}_{i,s}}\right)/ \left({DA^{\max}_{i,s} {\rm{-}} DA^{\min}_{i,s}}\right) \label{normalize_DA}\\
    RT^{norm}_{i,t,s} & = \left({RT_{i,t,s} {\rm{-}} DA^{ave}_{i,s}}\right) /\left({DA^{\max}_{i,s} {\rm{-}} DA^{\min}_{i,s}}\right). \label{normalize_RT}
\end{align}

\noindent
\underline{\textit{Step 4}}: For each $s$, $i$ and $t$, calculate normalized net load forecast error $\varepsilon^{norm}_{i,t,s}$ as 
\begin{align}
    \varepsilon^{norm}_{i,t,s} = DA^{norm}_{i,t,s}-RT^{norm}_{i,t,s}. \label{normalize_error}
\end{align}

The normalization in \textit{Step 3} is to ensure $\varepsilon^{norm}$ have a statistically significant pattern. Using $\varepsilon^{norm}$ as training data, the OA-cGAN will generate data ($\varepsilon^{gen}$) that follows the statistical characteristics of $\varepsilon^{norm}$, while maximizing the operating cost in the DC-OPF model.  
Note that the synthetic errors $\varepsilon^{gen}$ have to be transformed into a RT load value (``denormalized'') as:
\begin{align}
    d_{i,t,s} = \varepsilon^{gen}_{i,t,s} \left(DA^{\max}_{i,s} - DA^{\min}_{i,s} \right)+ DA_{i,t,s},  \label{denormalize}
\end{align}
where $d_{i,t,s}$ is the generated RT net load based on the generated forecast error and the real DA net load forecast.

\subsection{Training process}
\label{subsec:Gradient}
\begin{algorithm}[t]
    \small
    \SetAlgoLined
    \SetKwInOut{Input}{input}\SetKwInOut{Output}{output}
    \Input{$\{\varepsilon^{norm}_s\}_{s \in \set{S}_{train}}$\\
    }
    \Output{$\{\varepsilon^{gen}_s\}_{s \in \set{S}_{test}}$, $\theta_g$, $\theta_d$\\
    }
    \Begin{
      Initialize $\theta_g$ and $\theta_d$; $epoch \leftarrow 0$\\
      \While{$epoch < epoch^{\max}$ }{
      \For{$\set{B} \subset \set{S}_{train}$}
      {Input $\{\varepsilon^{norm}_s\}_{s \in \set{B}}$ to OA-cGAN;\\
      Obtain output of $G$ as $\{\varepsilon^{gen}_s\}_{s \in \set{B}}$;\\
      Calculate $\{loss_{D,s}\}_{s \in \set{B}}$ with $\{\varepsilon^{gen}_s\}_{s \in \set{B}}$;\\
      Update $\theta_d$ with $\{loss_{D,s}\}_{s \in \set{B}}$ using SGD; \\
      Calculate $\{loss_{G1,s}\}_{s \in \set{B}}$ with $\{\varepsilon^{gen}_s\}_{s \in \set{B}}$;\\
      Update $\theta_g$ with $\{loss_{G1,s}\}_{s \in \set{B}}$ using SGD; \\
      \For{$s \in \set{B}$}
      {Run DC-OPF \cref{DCOPF} based on $\varepsilon^{gen}_s$;\\
       Obtain $C^*_s$ based on \cref{C_star};\\
       Calculate $loss_{G2,s}$ based on \cref{objective_G};\\
      }
       Update $\theta_g$ with $\{loss_{G2,s}\}_{s \in \set{B}}$ using SGD based on the gradient in \cref{gradient_final} ; \\
      }
      $epoch \leftarrow epoch+1$
      }
      Obtain $\{\varepsilon^{gen}_s\}_{s \in \set{S}_{test}}$ based on fully-trained $G$\\
      \KwRet{$\{\varepsilon^{gen}_s\}_{s \in \set{S}_{test}}$, $\theta_g$, $\theta_d$}
     }
    \caption{Training process of OA-cGAN}
    \label{alg:OA_cGAN}
\end{algorithm}

Standard cGANs are trained using gradient-based methods.
In particular, the stochastic gradient descent (SGD), which uses an estimated gradient calculated from a randomly selected subset of the training data (so called ``mini-batch''), is the most common because it facilitates training over very large training data sets and exhibits superior convergence properties \cite{pmlr-v97-qian19b}.
Hence, the OA-cGAN can also rely on SGD to iteratively update (``train'') parameters $\theta_g$ and $\theta_d$. Nevertheless, the additional term of $loss_{G2}$ in \cref{objective_G}, which is related to the solutions to another optimization problem (i.e., DC-OPF), bring challenges to the direct use of SGD in the OA-cGAN. Thus, a suitable training method for the OA-cGAN needs to be designed.

Since the training process of $D$ is the same as it in traditional cGANs, it can be achieved by off-the-shelf functions that are readily implemented in many machine learning packages (e.g., TensorFlow, PyTorch, or Flux). Therefore, this section focuses on the parameter update method for the parameters of $G$ ($\theta_g$).
Since the two components of $loss_G$ are linearly additive, the gradient of $loss_G$ can be calculated by combining the gradient of $loss_{G1}$ and $loss_{G2}$. Thus, the resulting update rule for $\theta_g$ is:
\begin{align}
    \theta_g^{r+1} & {\rm{=}} \theta_g^r {\rm{-}} \frac {\alpha}{N_b} \sum\limits_{s \in \set{B}} \left( k \frac{{\partial loss_{G1,s}}}{{\partial \theta_g}} +  (1{\rm{-}}k) \frac{{\partial loss_{G2,s}}}{{\partial \theta_g}} \right), \label{SGD}
\end{align}
where $r$ denotes the training iteration, $\alpha$ is the learning rate, $\set{B}$ is the mini-batch of data from the training data set $\set{S}_{train}$, $N_b$ is the number of samples in mini-batch $\set{B}$, $loss_{G1,s}$ and $loss_{G2,s}$ are the losses of $G$ associated with sample $s$ in $\set{B}$. The progress of training in the SGD is measured by epochs, where one epoch means one complete pass of the training data set through the parameter update process. The gradient of $loss_{G1,s}$ is the same as the loss gradient of $G$ in traditional cGANs and, as for $D$, can be inferred using off-the-shelf implementations. 

Since $loss_{G2}$ does not explicitly contain $\theta_g$, to calculate the ${\partial loss_{G2,s}}/{\partial {\theta_g}}$, we need to find the relationship between $loss_{G2}$ and $\theta_g$ using intermediate variables. Analytically, for each sample $s$, $\theta_g$ decides the output of $G$ ($\{\varepsilon^{gen}_{i,t,s}\}_{i \in \set{I}, t \in \set{T}}$) which will affect the generated net load ($\{d_{i,t,s}\}_{i \in \set{I}, t \in \set{T}}$). Then, $\{d_{i,t,s}\}_{i \in \set{I}, t \in \set{T}}$ will affect the optimal output of generators ($\{p_{g,t,s}^*\}_{g \in \set{G}, t \in \set{T}}$) which will directly affect the optimal operating cost $C^*_s$ and thus $loss_{G2,s}$. Therefore, we can derive the gradient of $loss_{G2,s}$ using the chain rule as:
\begin{align}
    \frac{{\partial loss_{G2,s}}}{{\partial {\theta_g}}} = - \sum\limits_{t \in \set{T}} \sum\limits_{g \in \set{G}}  {\frac{{\partial {C_s^*}}} {\partial P_{g,t,s}^*} \sum\limits_{i \in \set{I}} \frac{\partial P_{g,t,s}^*}{\partial  d_{i,t,s}} \frac{\partial d_{i,t,s}}{\partial \varepsilon^{gen}_{i,t,s}} \frac{\partial  \varepsilon^{gen}_{i,t,s}}{\partial \theta_g}}. \label{gradient_loss2}
\end{align}
In the following, we derive each term in \cref{gradient_loss2}.

Term ${\partial {C_s^*}}/{\partial P_{g,t,s}^*}$ captures the marginal change of cost when changing the output of generator $g$ in the optimal solution of the DC-OPF at time $t$ in sample $s$. 
For generators with binding constraints \cref{DCOPF_Pmax} we have ${{\partial {C_s^*}}} /{\partial P_{g,t,s}^*} = 0$. 
For generators with non-binding constraints \cref{DCOPF_Pmax} (i.e., marginal generators) ${{\partial {C_s^*}}} /{\partial P_{g,t,s}^*} = \lambda_{i,t,s}/ {\delta_{scale}}$, where $\lambda_{i,t,s}$ is the locational marginal price at node $i$ and time $t$ in sample day $s$, i.e., the dual multiplier of \cref{DCOPF_power_balance}, and ${\delta_{scale}}$ is a constant value introduced in \cref{C_star}.
Note that $\lambda_{i,t,s}$ can be obtained directly from most numerical solvers after solving \cref{DCOPF}.
Therefore, we obtain:
\begin{align}
    \sum\limits_{g \in \set{G}} \frac{{\partial {C_s^*}}} {\partial P_{g,t,s}^*} =\sum\limits_{i \in \set{I}} \sum\limits_{g \in \set{G}_i} \frac{{\partial {C_s^*}}} {\partial P_{g,t,s}^*} = \sum\limits_{i \in \set{I}} \frac{\lambda_{i,t,s}}{{\delta_{scale}}}. \label{gradient_c_g}
\end{align}
Next, as per \cref{DCOPF_power_balance}, it follows:
\begin{align}
    \frac {\partial P_{g,t,s}^*}{\partial  d_{i,t,s}} = 1. \label{gradient_g_d}
\end{align}
Similarly, according to \cref{denormalize}, we obtain:
\begin{align}
    \frac{\partial d_{i,t,s}}{\partial \varepsilon^{gen}_{i,t,s}} = DA^{\max}_{i,s} - DA^{\min}_{i,s}. \label{gradient_d_e}
\end{align}
Finally, since $\varepsilon^{gen}_{i,t}$ is the output of $G$, ${\partial \varepsilon^{gen}_{i,t}}/{\partial \theta_g}$ can, again, be calculated by off-the-shelf implementations.

As a result, we can recast \cref{gradient_loss2} for SGD training as:
\begin{align}
    \frac{{\partial loss_{G2,s}}}{{\partial {\theta_g}}} = -\sum\limits_{t \in \set{T}} \sum\limits_{i \in \set{I}} \frac{\lambda_{i,t,s} (DA^{\max}_{i,s} - DA^{\min}_{i,s})}{{\delta_{scale}}} \frac{\partial  \varepsilon^{gen}_{i,t,s}}{\partial \theta_g}. \label{gradient_final}
\end{align}
Algorithm~\ref{alg:OA_cGAN} summarizes the  OA-cGAN training process.

\section{Numerical Experiments}
\label{sec:Case_study}

We apply the proposed OA-cGAN to protect the power system from the uncertain load through more accurate DA decisions.

\subsection{Reserve in the DA scheduling}
\label{ssec:reserve_testing}

To accommodate changes between the DA load forecast and the actual RT load, some generators need to provide reserves that are sufficient to offset the forecast error and that are deliverable through the transmission network, i.e., can be deployed without violating transmission constraints. 
At the same time, these reserves should be allocated in the least-cost manner. Assume a set of given forecast errors $\{\varepsilon_{i,t}\}_{i\in \set{I}, t\in \set{T}}$. The optimal reserve allocation can be calculated through the following modified DC-OPF formulation:
\allowdisplaybreaks
\begin{subequations}
\begin{align}
&\min_{\substack{\{P_{g,t}^{DA},r_{g,t}^{+},r_{g,t}^{-}\}_{g\in \set{G}, t\in \set{T}} \\  \{\theta_{i,t}, \bar \theta_{i,t}\}_{i\in \set{I}, t\in \set{T}}}}\ 
    C^{DA}=\sum\limits_{t\in \set{T}} \sum\limits_{g\in \set{G}} c^{DA}_{r} (r_{g,t}^{+} + r_{g,t}^{-})
     \nonumber \\ 
    & \hspace{1.5cm}+ \sum\limits_{t\in \set{T}} \sum\limits_{g\in \set{G}} {\left[c_{0g} + c_{1g}P_{g,t}^{DA} + c_{2g}({P_{g,t}^{DA}})^2 \right]} \label{DA_objective}\\
    &\text{s.t. } \forall t\in\set{T}:\\
    &\hspace{0.5cm} \sum\limits_{g\in \set{G}_i} P_{g,t}^{DA} {\rm{-}} \sum\limits_{j \in \mathcal{N}_i} B_{i,j} (\theta_{i,t} {\rm{-}} \theta_{j,t}) = DA_{i,t} \ \forall{i}\in\set{I} \label{DA_power_balance}\\
    &\hspace{0.5cm}\sum\limits_{g\in \set{G}_i}(r_{g,t}^{+} {\rm{-}} r_{g,t}^{-}) {\rm{-}} \sum\limits_{j \in \mathcal{N}_i} B_{i,j} (\bar \theta_{i,t} {\rm{-}} \bar \theta_{j,t}) {\rm{=}}  \varepsilon_{i,t} \ \forall{i}\in\set{I} \label{DA_reserve_balance} \\
    & \hspace{0.5cm} P_{g,t}^{DA} + r_{g,t}^{+} - r_{g,t}^{-} \le P_{g}^{\max} \quad \forall{g}\in\set{G} \label{DA_Pmax}\\
    & \hspace{0.5cm} -S_{i,j} \le B_{i,j} (\theta_{i,t} + \bar \theta_{i,t} - \theta_{j,t} - \bar \theta_{j,t}) \le S_{i,j}  \nonumber \\ 
    & \hspace{1cm}\forall{i}\in\set{I},\ \forall{j}\in \mathcal{N}_i \label{DA_power_flow}\\
    &\hspace{0.5cm} \theta_{ref,t} = 0, \ \bar \theta_{ref,t} = 0 \label{DA_ref} \\
    & \hspace{0.5cm} P_{g,t}^{DA} \ge 0, \ r_{g,t}^{+} \ge 0, \ r_{g,t}^{-} \ge 0 \quad \forall{g}\in\set{G},
\end{align}%
\label{DA_DCOPF}%
\end{subequations}%
\allowdisplaybreaks[0]%
where $P_{g,t}^{DA}$ is the day-ahead  output power of generator $g$ at time $t$, $r_{g,t}^{+}$ and $r_{g,t}^{-}$ are the upward and downward reserve provided by generator $g$ at time $t$, $\theta_{i,t}$ and $\bar \theta_{i,t}$ are the voltage angle at node $i$ and time $t$ considering only $DA_{i,t}$ and $\varepsilon_{i,t}$, respectively. Since the DC-OPF is a linear model, the voltage angles can be superimposed such that $\theta_{i,t} + \bar \theta_{i,t}$ is the voltage angle at node $i$ and time $t$ considering both $DA_{i,t}$ and $\varepsilon_{i,t}$.
Objective \cref{DA_objective} minimizes the DA operating cost ($C^{DA}$), which includes the power generation cost and the reserve provision cost. The price of day-ahead reserve provision $c_{r}^{DA}$ is set to \unit[20]{\$/MW} . Eqs.~\cref{DA_power_balance} and \cref{DA_reserve_balance} are the nodal power balance constraint, where  \cref{DA_power_balance} ensures the DA forecast net load is served by the active output power of generators, and \cref{DA_reserve_balance} ensures the DA net load forecast error is compensated for by the reserve provided by generators. Constraints \cref{DA_Pmax}-\cref{DA_ref} ensure deliverability of both scheduled generation $P_{g,t}^{DA}$ and reserves $r_{g,t}^{+}$, $r_{g,t}^{-}$.

If forecast error $\varepsilon$ was known exactly, then \cref{DA_DCOPF} would yield the optimal least-cost dispatch and reserve allocation. In practice, however, $\varepsilon$ is unknown and must be estimated. We can use the OA-cGAN to estimate forecast errors that are statistically credible but particularly ``stressful'', i.e., corresponding to a relatively large operating cost for the system.

\subsection{Balancing power in the RT scheduling}
\label{ssec:RT_balancing}

If the RT load is different from the DA load, then the generators need to provide balancing power during RT scheduling. The RT balancing power of generator $g$ at time $t$ can be calculated as $R_{g,t} = P^{RT}_{g,t} - R^{DA}_{g,t}$. Based on the relationship between $R_{g,t}$ and the DA scheduled reserve ($r_{g,t}^{+}$, $r_{g,t}^{-}$), we can divide the possible distribution interval of $R_{g,t}$ into two regions:
\begin{itemize}
    \item If $R_{g,t}$ is within the range of the DA scheduled reserve, i.e., $0 \le R_{g,t} \le r_{g,t}^{+}$ or $ 0 \ge R_{g,t} \ge -r_{g,t}^{-}$, then the balancing power is procured as part of the DA scheduling process. We denote this region of balancing power as \textit{Region I}. 
    \item If $R_{g,t}$ is beyond the range of the DA scheduled reserve, i.e., $R_{g,t} > r_{g,t}^{+}$ or $ -R_{g,t} \le r_{g,t}^{-}$, but is still within the technical limits of the generator, then the excess  balancing power, i.e., $R_{g,t}-r_{g,t}^{+}$ (if $R_{g,t}>0$) or $ -R_{g,t}- r_{g,t}^{-}$ (if $R_{g,t}<0$), is an impromptu emergency response and has not been planned as part of the DA scheduling process. We denote this region of balancing power as \textit{Region II}.
\end{itemize}
We set the cost $c_{r}^{\text{I}}$ and $c_{r}^{\text{II}}$ for providing balancing power in regions $R_{g,t}^{\text{I}}$ and $R_{g,t}^{\text{II}}$ to \unit[10]{\$/MW} and \unit[50]{\$/MW}, respectively.. 
The optimal RT balancing power can be calculated through the following formulation:
\allowdisplaybreaks
\begin{subequations}
\begin{align}
&\min_{\substack{\{P_{g,t}^{RT},R_{g,t}^{RT},R_{g,t}^{DA}\}_{g\in \set{G}, t\in \set{T}} \\  \{\theta_{i,t}\}_{i\in \set{I}, t\in \set{T}}}}\ 
    C^{RT} = \sum\limits_{t\in \set{T}} \sum\limits_{g\in \set{G}} (c^{\text{I}}_{r} R_{g,t}^{\text{I}} + c^{\text{II}}_{r} R_{g,t}^{\text{II}}) \label{RT_objective} \\ 
    &\text{s.t. } \forall t\in\set{T} \nonumber: \cref{DCOPF_power_flow}, \cref{DCOPF_ref} \\ 
    &\hspace{0.5cm} \sum\limits_{g\in \set{G}_i} P_{g,t}^{RT} - \sum\limits_{j \in \mathcal{N}_i} B_{i,j} (\theta_{i,t} - \theta_{j,t}) = RT_{i,t} \   
     \forall{i}\in\set{I} \label{RT_power_balance}\\
    & \hspace{0.5cm} R_{g,t}^{\text{I}} = \max \Big\{0,\ \min\{P_{g,t}^{RT} {\rm{-}} P_{g,t}^{DA},\ r_{g,t}^{+}\}, \nonumber \\
    & \hspace{2.5 cm} \min\{P_{g,t}^{DA} {\rm{-}} P_{g,t}^{RT},\ r_{g,t}^{-}\} \Big\} \quad \forall{g}\in\set{G} \label{RT_low_balancing}\\
    & \hspace{0.5cm} R_{g,t}^{\text{II}} = \max \big\{0, \ P_{g,t}^{RT} {\rm{-}} P_{g,t}^{DA} {\rm{-}} r_{g,t}^{+},\ P_{g,t}^{DA} {\rm{-}} P_{g,t}^{RT} {\rm{-}} r_{g,t}^{-} \big\} \nonumber \\  
    & \hspace{1cm} \forall{g}\in\set{G} \label{RT_high_balancing}\\
    & \hspace{0.5cm} 0 \le P_{g,t}^{RT} \le P_{g}^{\max} \quad \forall{g}\in\set{G} \label{RT_Pmax}
\end{align}%
\label{RT_DCOPF}%
\end{subequations}%
\allowdisplaybreaks[0]%
In the experiments below, we evaluate the generated forecast errors using the DA and RT operating costs ($C^{DA}$ and $C^{RT}$). Algorithm~\ref{alg:evaluation} summarizes the proposed evaluation method. 
\begin{algorithm}[t]
    \small
    \SetAlgoLined
    \SetKwInOut{Input}{input}\SetKwInOut{Output}{output}
    \Input{$\{DA_s, RT_s, \varepsilon_s\}_{s \in \set{S}_{test} }$;\\
    number of samples in $\set{S}_{test}$ ($N_{test}$)
    }
    \Output {%$C_{total}$, $I^+$, $I^-$
    DA operating cost ($C^{DA}$);\\
    RT operating cost ($C^{RT}$)
    }
    \Begin{
      $C^{DA} \leftarrow 0$, $C^{RT} \leftarrow 0$; \\
      \For{$s \in \set{S}_{test}$}
      {Run day-ahead DC-OPF \cref{DA_DCOPF} with $DA_s$ and $\varepsilon_s$, obtain $C^{DA}_s$, $\{P_{g,t,s}^{DA}, r_{g,t,s}^{+}, r_{g,t,s}^{-}\}_{ g \in\set{G}, \ t\in\set{T}}$;\\
      $C^{DA} \leftarrow C^{DA} + C^{DA}_s$;\\
      Run real-time DC-OPF \cref{RT_DCOPF} with $RT_s$ and $\{P_{g,t,s}^{DA}, r_{g,t,s}^{+}, r_{g,t,s}^{-}\}_{ g \in\set{G}, \ t\in\set{T}}$, obtain $C^{RT}_s$;\\ 
      $C^{RT} \leftarrow C^{RT} + C^{RT}_s$;\\
      }
      \KwRet{$C^{DA}$, $C^{RT}$}
     }
    \caption{Evaluation of Given Error $\varepsilon$}
    \label{alg:evaluation}
\end{algorithm}

\subsection{Test system and data}
\label{subsec:System}

\begin{figure}[!t]
    \centering
    \includegraphics[width=0.7\linewidth]{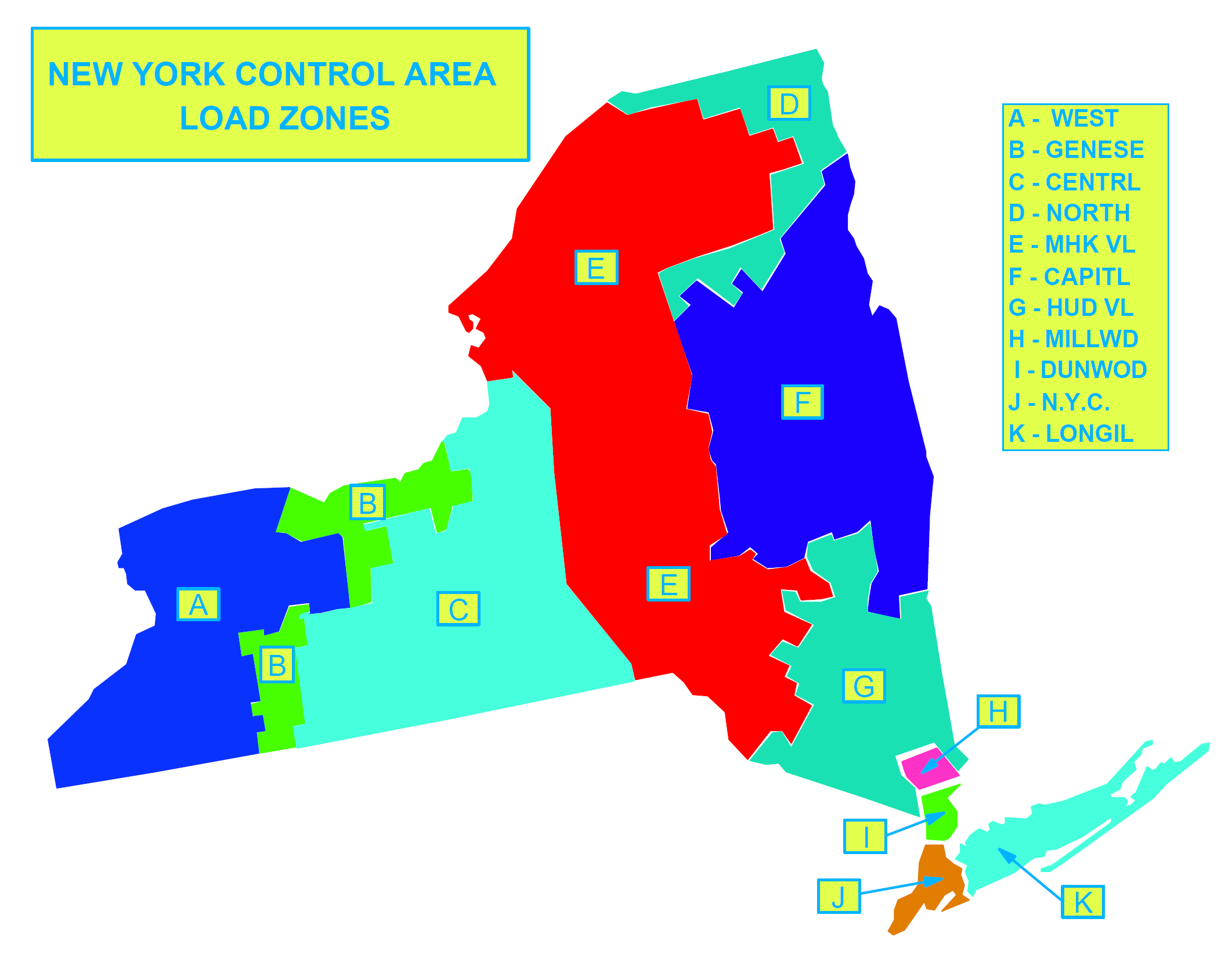}
    \caption{11-zone representation of the NYISO system \cite{NYISO_zone_map}.}
    \label{fig:zones}
\end{figure}

We conduct our numerical experiments using a zonal representation of the New York Independent System Operator (NYISO) system, as shown in Fig.~\ref{fig:zones}. 
Following the NYISO market structure, the full system is aggregated into an 11-zone system. (We note that this 11-zone representation is used in real-world operations for computing locational marginal prices for load charges).
The hourly DA net load forecasts and actual RT net loads for each zone are available from NYISO in \cite{NYISO_load}.
The system is populated with 362 generators and 33 wind farms, whose locations and parameters have been estimated from publicly available data bases \cite{NYISO_Gold_Book, Wind_Turbine_Database}.
All computations were carried out in Julia v1.5 \cite{Julia-2017}. The neural networks in the OA-cGAN were built and trained using the Flux package \cite{innes2018}, and the DC-OPF problems were implemented in JuMP \cite{DunningHuchetteLubin2017} and solved using the Gurobi solver \cite{gurobi}. 
All experiments were performed on a standard PC workstation with an Intel i9 processor and 16 GB RAM. The training time of the OA-cGAN ($k \neq 1$) for each epoch was around 5 minutes, while the training time of a traditional cGAN ($k=1$) for each epoch was around 1 minute.
Our implementation and data is publicly available at \cite{oacgan_code}.

\begin{figure}[!t]
    \centering
    \includegraphics[width=1\linewidth]{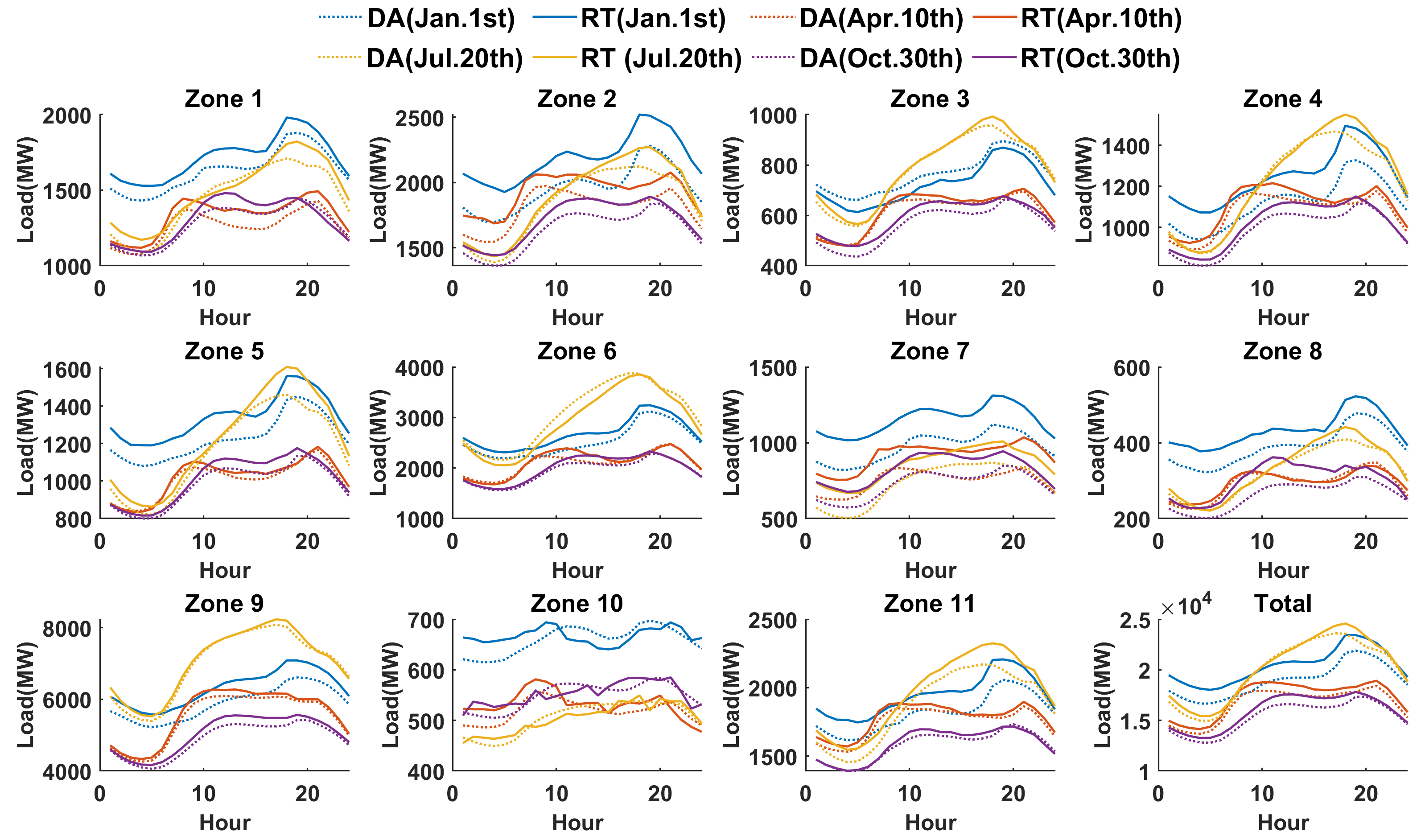}
    \caption{Day-ahead and real-time loads in 11  NYISO zones. ``Total'' shows the sum over all 11 zones.}
    \label{fig:real_load}
\end{figure}

We obtained the hourly DA and RT net load from January 1st, 2018 to January 4th, 2021 (1100 days in total) from  NYISO  and randomly split the data in 1050 days for training the OA-cGAN, as described in Sections~\ref{subsec:Training} and \ref{subsec:Gradient}, and 50 days for testing as described in Section~\ref{ssec:reserve_testing}.
Fig.~\ref{fig:real_load} shows the DA and RT load profiles for four selected days (Jan. 1st, Apr. 10th, Jul. 20th, Oct. 30th 2018) in each zone drawn from each quarter of the year.
It can be seen that for each zone and each quarter the forecast errors exhibit distinct seasonal characteristics. Thus, we decide to use quarters at labels in the OA-cGAN training, i.e., the load in the first (Jan.--Mar.), second (Apr.--Jun.), third (Jul.--Sep.) and fourth (Oct.--Dec.) quarter are labeled as 0, 1, 2, and 3 respectively. 
Note that this labeling system is used in this paper for the simplicity of illustration. To generate errors with more specific properties, one can use more complicated labeling systems, which include more information of the target day, such as the daily temperature or the precipitation.

\begin{figure}[!t]
   \centering
    \includegraphics[width=1\linewidth]{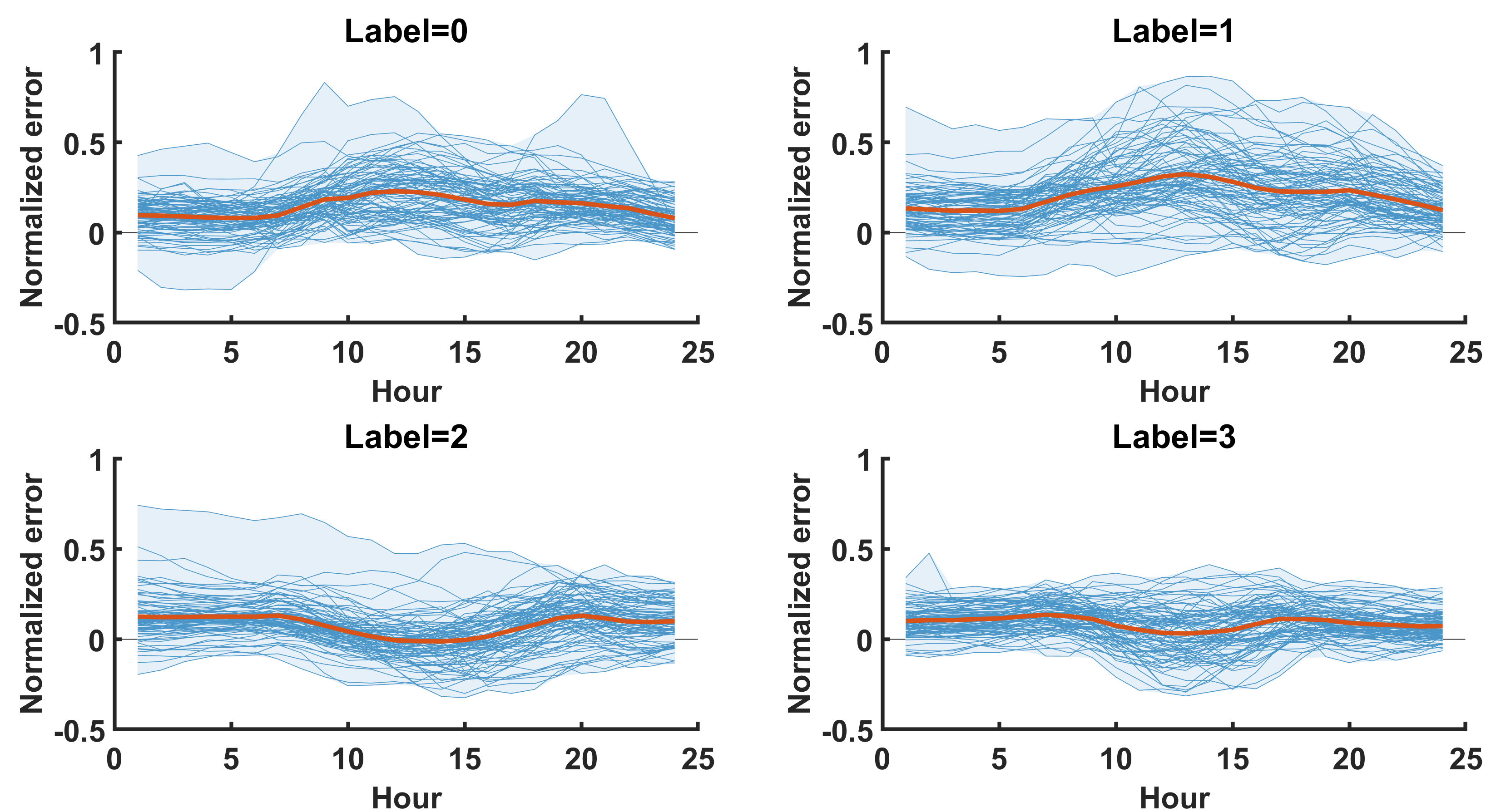}
    \caption{Historical normalized errors in Zone 1 for each label.}
    \label{fig:error_zone 1}
\end{figure}

\begin{figure}[!t]
    \centering
    \includegraphics[width=1\linewidth]{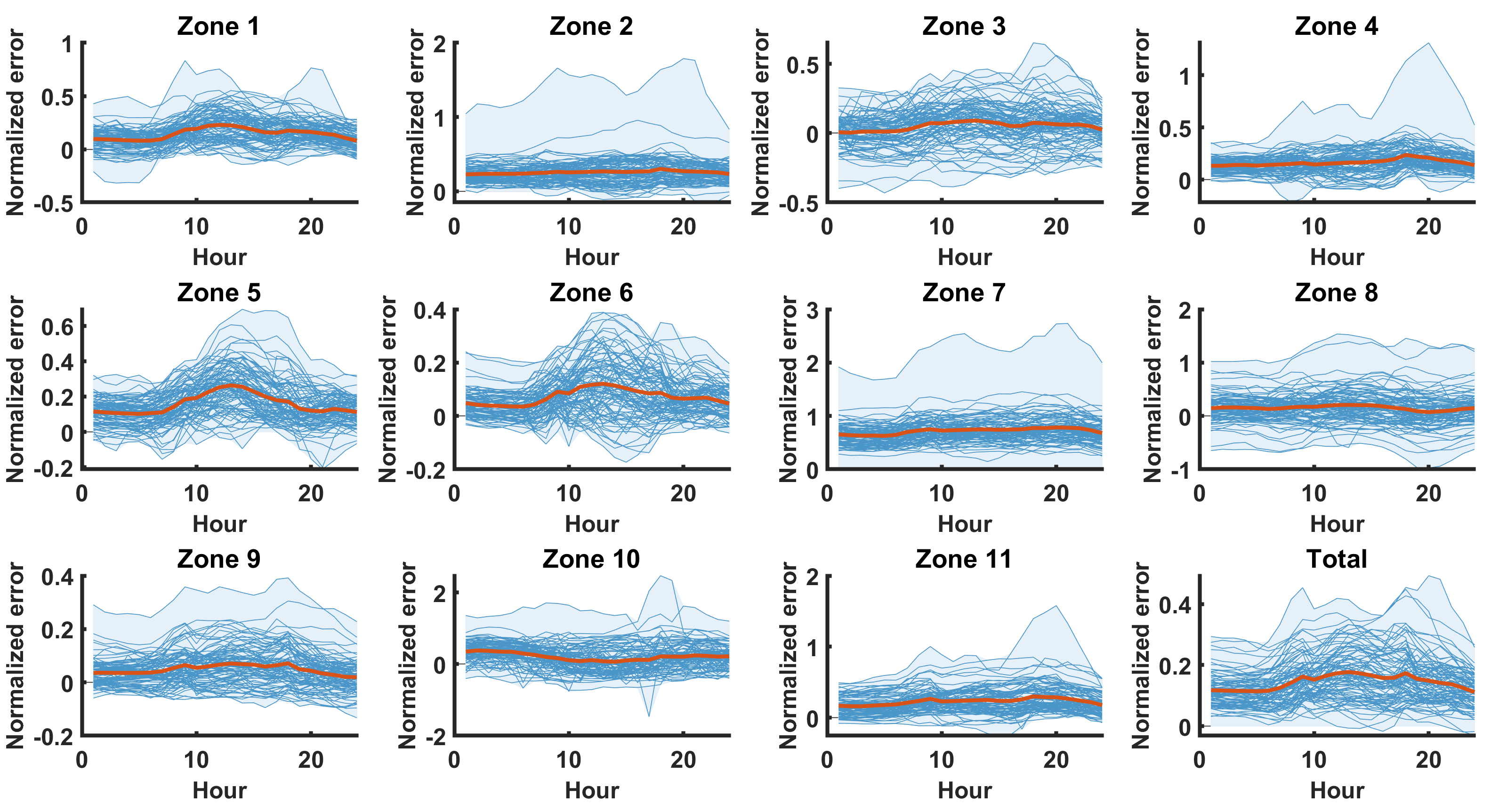}
    \caption{Historical normalized errors in 11 NYISO zones and the total sum for Label=0.}
    \label{fig:error_12_zones}
\end{figure}

The normalized errors in Zone 1 with different labels are shown in Fig.~\ref{fig:error_zone 1}, and the normalized errors in the 11 zones and the whole NYISO system with the same label (label=0) are shown in Fig.~\ref{fig:error_12_zones}. The blue lines in Fig.~\ref{fig:error_zone 1} and Fig.~\ref{fig:error_12_zones} are the real normalized errors in year 2018, the red line in the middle of each sub-figure is the average of the blue lines in the same sub-figure, and the light blue areas indicate the possible distribution area of the errors according to historical data.

The rationality of the labeling system in this case study is further illustrated in Fig.~\ref{fig:error_zone 1}. Note that the characteristics of errors with different labels are significantly different from the following two aspects. First, the average errors with different labels have different shapes. For example, the average error curves with labels 0 and 1 are upward protruding, while the curves with labels 2 and 3 are downward protruding. Second, the width of the distribution areas of errors with different labels are different. For example, the distribution area of errors with label 1 is noticeably wider than  with label 3. Thus, with this labeling system, the characters of errors can be distinguished.

Fig.~\ref{fig:error_12_zones} displays differences among the normalized errors in 11 zones. For example, in Zone 3, the errors approximately evenly distribute between $-0.5$ and $0.5$ and have an average value close to 0, while in Zone 3, all the errors are positive and the maximum error is around 2.

\subsection{Training Results}
\label{subsec:Training_Results}

We test the OA-cGAN using seven different values of $k$ between $0.5$ and $1$. Recall that $k$ determines the weight between the two objectives of $G$ in Eq.~\cref{objective_G}, i.e., the objective of $G$ is to generate statistically credible errors (minimizing $loss_{G1}$) that are also operational-adversarial (minimizing $loss_{G2}$). The greater the $k$ is, the more important the first objective is.

\begin{figure}[!t]
    \centering
    \includegraphics[width=1\linewidth]{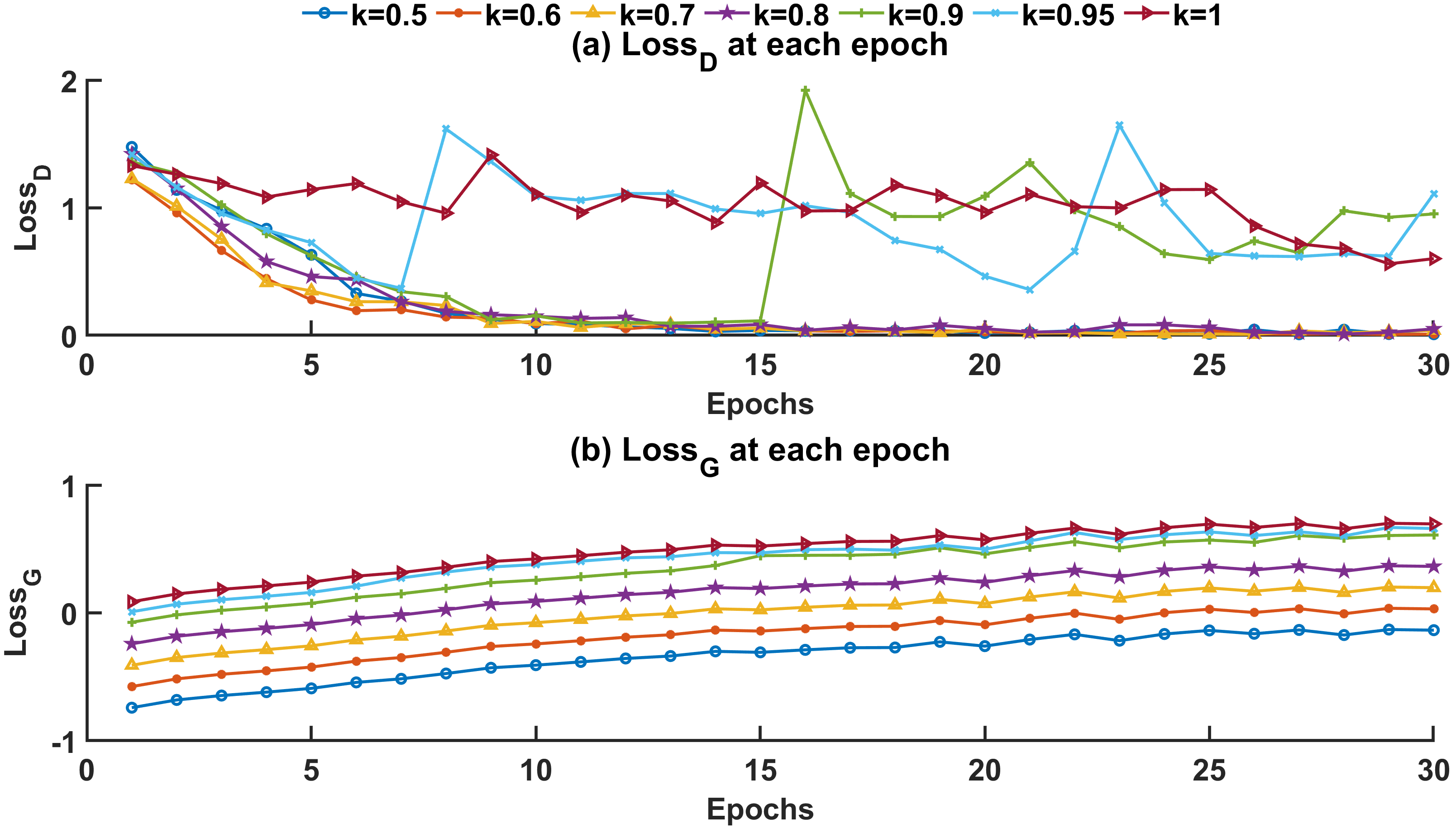}
    \caption{Values of loss function terms $loss_{D}$ (a) and $loss_{G}$ (b) during the training process.}
    \label{fig:loss_G_D}
\end{figure}

Both $G$ and $D$ in the OA-cGAN in this numerical experiment are three-layer convolutional neural networks which use rectified linear units (ReLU) as the activation function. The size of the mini-batch during the training process is 100, and the reference cost for scaling in \cref{C_star} are $\delta_{shift}=2 \cdot 10^8$ and $\delta_{scale}=8\cdot 10^5$. The loss of $G$ and $D$ during the first 30 epochs of the training process is shown in Fig.~\ref{fig:loss_G_D}. According to Fig.~\ref{fig:loss_G_D}(a), the overall trend of $loss_D$ during training process is rapidly decreasing at the first 10 epochs and then gradually stabilizes. On the contrary, according to Fig.~\ref{fig:loss_G_D} (b), the overall trend of $loss_G$ during the training process is slowly increasing at first and then gradually stabilizes.

Based on the pattern of each curve in Fig.~\ref{fig:loss_G_D}, we can divide the seven cases with the seven different values of $k$ into two groups, i.e., the three cases when $k \ge 0.9$ are in one group and the four cases when $k \le 0.8$ are in another group. When $k \ge 0.9$, $loss_D$ converges to 1, indicating that the discriminator cannot identify whether the input data are original or generated very well; while when $k \le 0.8$, $loss_D$ converges to $0$, indicating that the discriminator can almost completely distinguish the generated and the original data and the credibility of the generated data is poor. With $k=0.9$, $0.95$ and $1$, $loss_G$ converges to three close positive values between $0$ and $1$. However, if $k \le 0.8$, $loss_G$ converges to four equally spaced negative values.

\begin{figure}[!t]
    \centering
    \includegraphics[width=1\linewidth]{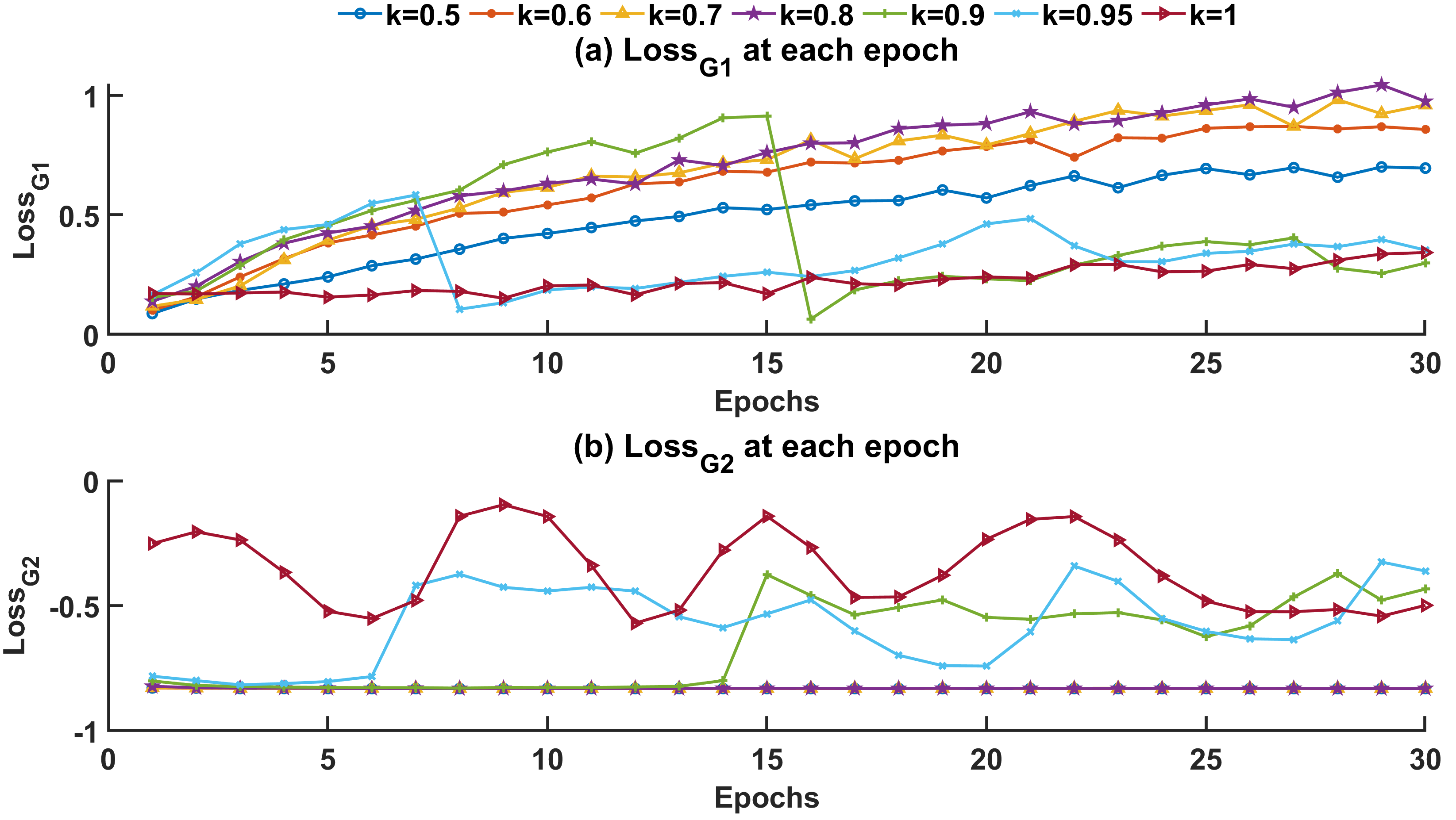}
    \caption{Values of loss function terms $loss_{G1}$ (a) and $loss_{G2}$ (b) during the  training process.}
    \label{fig:G_loss_epoch}
\end{figure}

To further explain the results of $loss_G$, we plot the value of $loss_{G1}$ and $loss_{G2}$ during the training process separately in Fig.~\ref{fig:G_loss_epoch}(a) and (b). According to Fig.~\ref{fig:G_loss_epoch}(b), when $k=1$, $loss_{G2}$ oscillates between $-0.5$ and $0$. When $k<1$, $loss_{G2}$ at the beginning of the training process is around $-0.7$, and only in the cases when $k \ge 0.9$, $loss_{G2}$ deviates from the initial value during the training process and starts to oscillate. When $k \le 0.8$, $loss_{G2}$ will not change during the training process.

Moreover, in the cases when $k=0.9$ (or $0.95$), there is an obvious turning point at epoch 15 (or epoch 7) on the curves of $loss_{G1}$, $loss_{G2}$, and $loss_D$, but there is no turning point on the curve of $loss_{G}$. These turning points reflect the changes in the relative influence of the two objective of $G$ during the training process. For example, when $k=0.9$, the objective of minimizing $loss_{G2}$ controls the training of $G$ before epoch 15. Thus, during this period, $loss_{G2}$ remains at a low level, while $loss_{G1}$ keeps increasing and $loss_{D}$ keeps decreasing because $D$ can recognize generated data better and better. After the turning point, the influence of minimizing $loss_{G1}$ exceeds the influence of minimizing $loss_{G2}$, so $loss_{G1}$ starts to decrease and $loss_{G2}$ starts to increase. In the cases when $k \le 0.8$, $loss_{G1}$ increases for the whole training process, indicating that minimizing $loss_{G2}$ dominates minimizing $loss_{G1}$. 

\subsection{Testing Results}
\label{subsec:Testing_Results}

In this section, we present the testing results of the fully trained OA-cGAN. According to the training results in Section \ref{subsec:Training_Results}, the generated data of OA-cGAN when $k \le 0.8$ have similar characteristics. Thus, we will only study the testing results when $k$ is equal to $0.8$, $0.9$ or $1$. 

\begin{figure}[!t]
    \centering
    \includegraphics[width=1\linewidth]{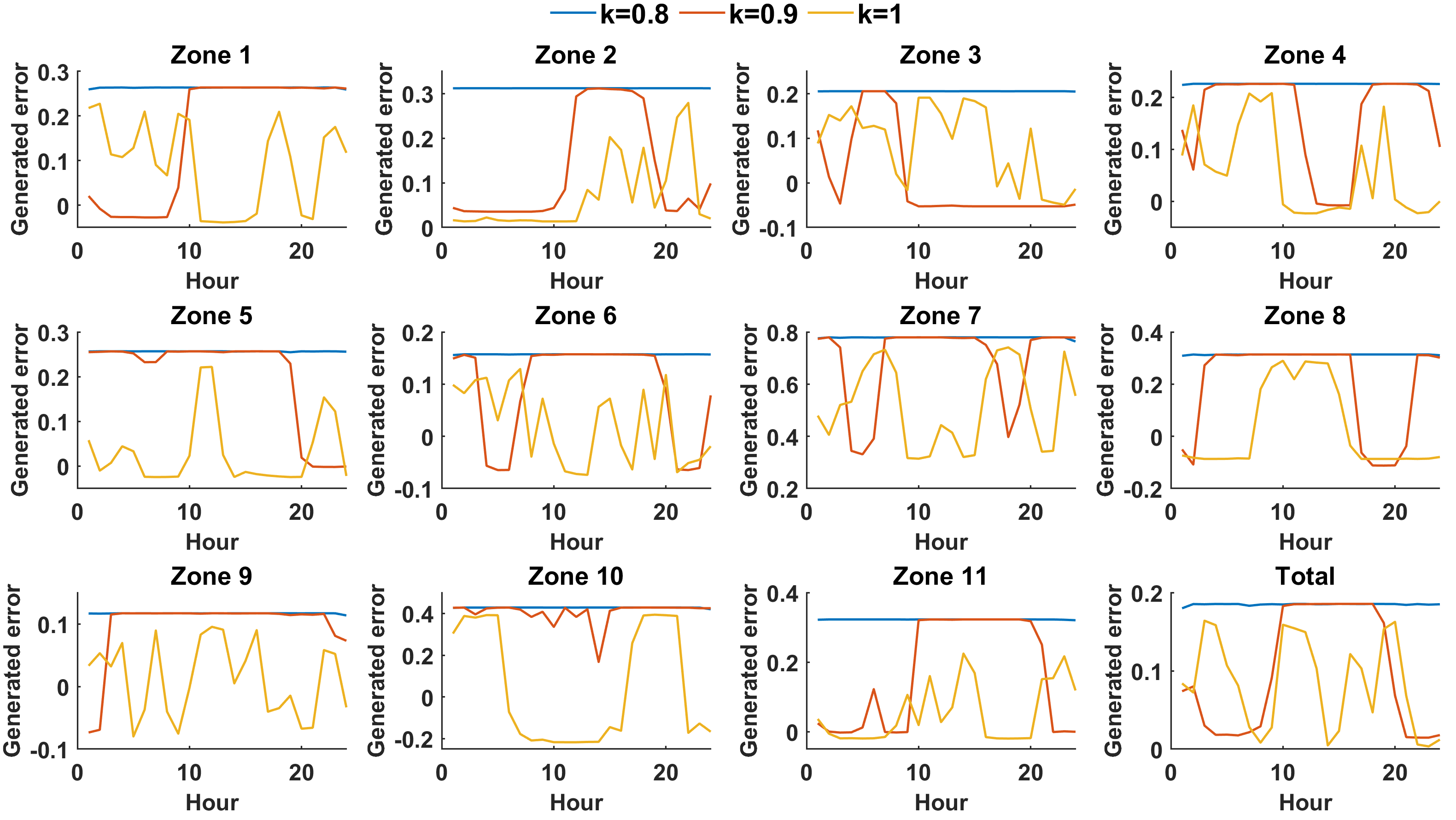}
    \caption{Selected generated errors for the fully trained OA-cGAN for all zones and label=0.}
    \label{fig:generated_error}
\end{figure}

Errors generated by the trained OA-cGAN are shown in Fig.~\ref{fig:generated_error}. We notice that the greater the $k$, the greater the variance of the generated errors. Specifically, the generated errors for 11 zones and the whole system when $k=0.8$ are always straight lines, which corresponds to the most costly cases in each zone. Note that the maximum values of the generated errors in 11 zones are different, which corresponds to the historical error distributions of each zone shown in Fig.~\ref{fig:error_12_zones}.

Then, we will evaluate the performance of the generated errors using Algorithm~\ref{alg:evaluation}. We compare the evaluation results of the generated errors ($\varepsilon^{gen}$) by the OA-cGAN and the robust errors ($\varepsilon^{robust}$), which are assumed to be proportional to the real DA load ($DA$) as:
\begin{align}
    \varepsilon^{robust}_{i,t} = r DA_{i,t}, \quad \forall{i}\in\set{I},\ \forall{t}\in\set{T} \label{robust_error}
\end{align}
where $r$ is the level of robustness.

Table~\ref{tab:evaluation} summarizes the performance for each of the seven cases.
In Case 1, we do assume no forecast error between the DA and RT stages, so no reserve will be deployed in RT operation. In Cases 2-4, we consider errors generated by the OA-cGAN with different values of $k$. In Cases 5-7, we consider robust errors generated as \cref{robust_error} with different values of $r$. It can be seen from Table~\ref{tab:evaluation} that from Case 1 to Case 7, the DA operating cost $C^{DA}$ monotonically increases, while the RT operating cost $C^{RT}$ monotonically decreases. 
This observation correlates to the errors used in DA scheduling as larger errors lead to more reserve procurement at the DA stage and less emergency RT balancing power in the RT stage. As a result, Case 4 achieves the lowest overall cost $C^{total} = C^{DA}+ C^{RT}$.

\begin{table}[!t]
  \centering
  \caption{Evaluation results of each testing case (in million \$)}
    \begin{tabular}{ccccc}
    \toprule
    Case No. & Error type & $C^{DA}$  & $C^{RT}$ & $C^{total}$\\
    \midrule
    1 & No error  & 500.22  & 358.86 & 859.07 \\
    \midrule
    2 & Generated error ($k=1$) & 505.33 & 355.98 & 861.31 \\
    3 & Generated error ($k=0.9$) & 508.91 & 327.62 & 836.52  \\
    4 & Generated error ($k=0.8$) & 511.73 &309.80 	& 821.53 \\
    \midrule
    5 & Robust error ($r=0.1$) & 542.27 & 283.42 & 825.69 \\
    6 & Robust error ($r=0.3$) & 750.83 & 115.87 & 866.70 \\ 
    7 & Robust error ($r=0.5$) & 1727.50 & 3.06 & 1730.56  \\
    \bottomrule
    \end{tabular}%
  \label{tab:evaluation}
\end{table}

\section{Conclusion}
We developed an operation-adversarial conditional generative adversarial network that internalizes a DC optimal power flow model to generate statistically credible, stressed net load scenarios that are stressful, where the degree of stress is measured in the system operating  cost. 
The numerical experiments based on a real-world NYISO 11-zone system demonstrated that the net load forecast errors produced by the OA-cGAN lead to generator dispatch and reserve allocations that are more cost effective than robust benchmarks. The proposed OA-cGAN model could become an extension of the current power system scheduling procedure and it can also be used to generate stressful samples of other uncertain parameters, such as wind and solar power. 

\bibliographystyle{IEEEtran}
\bibliography{literature}

\end{document}